\documentclass[%
 aip,
 amsmath,amssymb,
 reprint,%
]{revtex4-1}

\usepackage{graphicx}
\usepackage{dcolumn}
\usepackage{bm}

\usepackage[utf8]{inputenc}
\usepackage[T1]{fontenc}
\usepackage{mathptmx}
\usepackage{etoolbox}

\makeatletter
\def\@email#1#2{%
 \endgroup
 \patchcmd{\titleblock@produce}
  {\frontmatter@RRAPformat}
  {\frontmatter@RRAPformat{\produce@RRAP{*#1\href{mailto:#2}{#2}}}\frontmatter@RRAPformat}
  {}{}
}%
\makeatother
\begin{document}

\preprint{AIP/123-QED}

\title{Comment on: \textit{In vitro} prediction of the lower/upper-critical biofluid flow choking index and \textit{in vivo} demonstration of flow choking in the stenosis artery of the animal with air embolism}
\author{T. Podgorski}
 \email{thomas.podgorski@univ-grenoble-alpes.fr}
\affiliation{ 
Universit\'e Grenoble Alpes, CNRS, Grenoble INP, LRP, F-38000 Grenoble, France
}%


\date{\today}

\begin{abstract}
Sanal Kumar \textit{et al.} \cite{kumarPHF22a} present a model of transonic compressible flows based on ideal gas theory that is irrelevant to biofluid flow and there are flaws in the general reasoning. In addition, the experimental attempts do not show any evidence of supersonic flow and do not provide any support for the flawed theory.  
\end{abstract}

\maketitle

In a recent paper, Sanal Kumar \textit{et al.} \cite{kumarPHF22a} propose predictions of a flow choking index that is supposed to be relevant to the understanding of various vascular strokes and diseases which in their opinion are supposed to be triggered when a critical Mach number is reached. They also provide their own interpretation of issues related to air embolisms and describe experimental and numerical results which are claimed to be supporting their theory. However, besides the fact that the logical structure of the article is fuzzy, there are several major issues with this work, from the framework and applicability of the model, to logical fallacies in its development and conclusions as well as false statements on blood properties and irrelevant or non-physical experimental results. Only the main and most salient controversial points are discussed here. Note that some of these criticisms also hold for their previous paper in this journal\cite{kumarPHF22b} as it is based on the exact same erroneous model.

\section{Relevance of the model}

The main model equations (Eqs (1-5)) which were derived in one of the authors' previous publications \cite{kumar18} are simply based on a mass conservation equation and the classic equation of state and expression of the speed of sound for an \textit{ideal gas}. This assumption on which the model is based is already a major flaw since liquids, biofluids including blood can certainly not be considered as ideal gases (if it were so, following the well known $P V =n R T$ law, our body volume would double when going from sea level to Himalayan mountains).

In other words, even considering that (in the strict sense) any material, solid or fluid, is compressible, the ideal gas equation is not even remotely an approximation of their behavior. Blood being mainly composed of water, its compressibility is close to that of water, a well-tabulated property which is around $4.4 \times 10^{-10}$~Pa$^{-1}$ at body temperature and around atmospheric pressure\cite{fine73}, with only little variations with pressure and temperature. This is $10^5$ times lower that the compressibility of ideal gases at atmospheric pressure. This fully justifies that blood is safely  considered nearly incompressible for all practical flow conditions with solid proof from a plethora of theoretical, numerical and experimental studies available in the literature. Equations (1-5) are therefore irrelevant to biofluids and only apply to ideal gases.

Now, even supposing that these equations could be modified by replacing the ideal gas law by the equation of state for a real fluid, the general approach which consists in looking for critical Mach numbers (i.e. suspecting that supersonic flows occur in the cardiovascular system) is bound to fail because the well documented range of blood flow velocities is between 1 mm/s (capillaries) and less than 1 m/s (arteries) everywhere in the circulatory system\cite{Robertson2008,BFLV} while on the other hand the speed of sound in blood, which is also well documented in  textbooks\cite{Duck1990} and widely used in medical ultrasonic techniques is close to that of water, that is about 1500 m/s. In other words, the Mach number in blood flows is always less than $10^{-3}$, meaning that the "Sanal flow choking" theory presented in the commented paper (and other similar papers by the same group of authors), even if it were to be corrected for real fluids, is irrelevant to cardiovascular flows. 

Following this clarification, it is clear that none of the \textit{in silico} results presented in Figures 7-11 of the paper represent realistic blood flow in arteries. No numerical simulation of blood flow in arteries using relevant values of physical parameters (viscosity, density) and boundary conditions (flow rates, pressures, velocities, channel dimensions), even including turbulence modeling \cite{lui2020} would lead to the flow patterns shown in Figures 7-11 where the Mach number goes up to 2. Unfortunately, none of the parameter values, boundary conditions and length scales used in these simulations are given in the article (the axial distance axis has no units in any of these figures).
Only the video linked to Fig. 8 shows a scale bar, from which we note that their simulated artery has an irrelevant diameter between 10 cm and 20 cm. In addition, the inlet Mach number in these figures also has completely unrealistic values of order 1, meaning that the authors assume that blood velocity has nonsensical values of the order of kilometers per second when entering the depicted artery section. Stunningly, some of these figures are \textit{the exact same} as the ones they show in other papers about aircraft or rocket aerodynamics, for instance Figure 7, which according to the caption shows numerical simulations of blood flow in an artery, is exactly the same as Figure 1 in a previous paper on aircraft dynamics\cite{kumarPHF21}, and has also been used in another previous paper where it is supposed to apply to different cases\cite{kumarPHF22b}. Therefore, these simulation results which were performed with unrealistic boundary conditions and unrealistic parameter values must be totally discarded with regard to their applicability to blood flows. 

As an additional note, model equations  are based on circular reasoning\cite{kumar18}: they assume that a boundary layer develops in the channel (through unspecified mechanisms), which effectively reduces the cross-section of the flow, which because of mass conservation leads to an increase of the flow velocity, leading to equation 5 which is nothing more than a reformulation of the trivial relationship between the local hydraulic diameter and local velocity in a channel imposed by mass conservation. This triviality can obviously not predict the existence nor  the thickness of this boundary layer as a function of distance along the flow axis  since it is a single equation relating two unknowns (the so-called boundary layer thickness and the local average velocity). The specific heat capacities of water (blood's main constituent) are $C_p=4.18$ kJ/(kg.K) and $C_p=4.09$ kJ/(kg.K) at $T=37^\circ$C as can be found in any handbook\cite{Raznjevic95} (values are similar for blood\cite{Mcintosh2010}), the heat capacity ratio is then $\gamma=1.02$. It is indeed always close to 1 in all physiologically relevant cases contrary to what the authors want us to believe in the introduction (it only goes up to $1.03$ at $T=42^\circ$C,  lethal temperature). Therefore the trivial mass conservation equation (Eq. 5a) almost reduces to $U d^2 \simeq$ constant and no "choking" or supersonic velocity is expected with realistic inlet velocities, pressures and channel diameters, even in the most pathological cases. Finally, equation 1 cannot represent systolic to diastolic pressure ratios since their modeling does not involve any time dependency or transient dynamics but only steady flow: it is an illicit use and combination of classic equations of compressible fluid theory taken out of context.

\section{Flawed experiments}

In an attempt to support their theory (which for the reasons stated above is irrelevant to blood and is based on flawed and circular reasoning), the authors describe a confusing ensemble of experimental measurements that do not provide much valuable or relevant results and do not support their claims.

\begin{itemize}
    \item The measurements of the heat capacity ratio (HCR) of blood  provided in Table 1 are completely unphysical, ranging from 5 to 118. No material even the most exotic one exhibits such values and even for ideal gases, the equipartition theorem states that the maximum value is $5/3\simeq 1.667$ for monoatomic gas. Values shown in Table 1 are therefore wrong. 
    \item A confusing sentence in the methodology section says \textit{"We observed that around 60-85$^\circ$C, all the blood samples of human being boiloff in a non-linear fashion"}. Blood, which is made mainly of water, does not boil at these temperatures.
    \item Similarly, in section II, they say \textit{"It is crystal clear from the case report of Razavi et al. that the COVID-19 patient suffers from gas embolism because the patient's temperature exceeds 37.5 $^\circ$ C"}. This entails two serious problems, the first one being that the cited publication (reference 8 in their publication list) does not make any mention of gas embolism in the reported clinical case, it is therefore pure speculation from the authors. 
    Then, they seem to imply in this sentence (and at other instances in the paper) that blood starts boiling or evaporating at temperatures above 37.5 $^\circ$C, Which, had it been true, would not have allowed the development of life on Earth as we know it.  Fortunately, the boiling point of blood (mainly composed of water) is close to 100$^\circ$C at atmospheric pressure. Gas embolisms occur only in two situations, either following accidental injection of gas bubbles in the circulation (i.e. during surgery) or due to degassing of dissolved gases (N$_2$, CO$_2$) following sudden and large decompression (i.e. underwater divers ascending too quickly to the surface). The reported COVID-19 case is obviously not concerned.
    \item The \emph{in vivo} experiment of Fig. 12 is perplexing: they injected high-pressure air into a rat's artery to induce air embolism. It is hard to understand what they were trying to prove: no quantitative pressure and velocity measurements were made and the flow rate imposed by the syringe pump is not specified either. Despite the authors' claim, the numerical simulations of Fig. 10 and 11 are not correlated to the experiment at all since no parameter value corresponds to experimental ones.  This cannot constitute a proof of the claims of the paper.
    \item Figures 17, 18 and 19 
    only show compressed air being blown through a plastic tube but do not demonstrate anything either qualitatively (no specific observations are described) or quantitatively since no quantitative measurements are made in these experiments. Besides, the connection between these situations and blood flow in arteries is virtually inexistent. And there is not even a demonstration of supersonic flow in these experiments.
    \item More generally the focus on air embolisms in the second half of the paper highlights a logical fallacy in the study: initially, the authors discuss blood flow, 
    attempt to measure its calorimetric properties in order to 
    support their theory, but in the end they claim to make experimental demonstrations with air. 
\end{itemize}

In conclusion, the experimental investigations performed by the authors and comments based on other studies from the literature do not support the claims derived from the model, which itself is based on wrong assumptions.

\nocite{*}
\bibliography{comment}

\end{document}